# The BlueNetwork Concept

*Reza FARRAHI MOGHADDAM, Fereydoun FARRAHI MOGHADDAM and Mohamed CHERIET*

Synchromedia Lab, ETS, University of Quebec
1100 Notre-Dame west, Montreal, H3C 1K3 Canada

## Table of Contents

## Summary

Here, we propose a social network, BlueNetwork, which directly targets behavioral engineering of the humans behavior in order to regulate and adjust it through human-human and human-machine interactions toward eliminating unnecessary and extra demand on the machines and the earth resources. This approach consequently reduces harming human footprints such as energy and water consumption, and GHG emissions. The proposed network provides human interfaces on various platforms (web and cellphone), and includes two artificial intelligence software units which respectively learn: 1) common-sense behavioral schedule of a healthy human and 2) constructive communication between a healthy human and another human. In order to be specific, volunteer declaration that a human is not obese will be used as the criterion to differentiate healthy humans (Blue humans) from the others. The sub-targets are as follows:

1. Providing directed and selective interaction between healthy and other humans over various platforms.
2. Developing software to learn the behavior (schedule) of healthy humans.
3. Developing software to learn effective communication between healthy and other humans toward achieving better life-style for them.
4. Providing assistance to healthy humans in order to keep them on shape and on their healthy schedule.

## Motivation

Most of energy efficiency and carbon reduction initiatives and concepts attempt to regulate and optimize machines behavior, and therefore, human behavior itself is left neglected. Although most of energy and resource consumption is the result of machines functioning and behavior (including domesticated animals such as cows), these behaviors themselves are actually in answer to humans demands and needs, and therefore, can be considered the indirect results of humans behavior. Resolving the source of problems, i.e., the unhealthy human behavior, not only reduces these footprints including energy and water consumption, and GHG emissions, it also helps increasing the quality of life in society.

Here, we propose an approach which focuses on adjusting humans behavior toward eliminating unnecessary demand on the machines that consequently lowers the consumption. This goal is achieved by creating a social environment in which directed and selective interactions help humans to adjust to healthier behavior. The solution consists of human-friendly interfaces and also artificial intelligence software in order to learn and emulate human interactions.

## Definitions

In this section, some terms used in the rest of paper are defined.

1. BMI (Body Mass Index): BMI = 703 mass (lb)/ (height (in))$^2$ = mass (kg)/ (height (m))$^2$.
2. Obesity: Volunteer declaration by an individual will be used to assign him obese on not-obese. However, in general, a BMI > 30, or about 30 lbs overweight for 5'4"person can be a sign of obesity.
3. Safe (Blue) human: A human being that himself observes that he is in a safe health condition. This is confirmed by his declaration of not being obese.
4. At-risk (Red) human: A human being that himself observes that, because of being obese, he is at a risky health condition. This is confirmed by his declaration.
5. Purple agent: An artificial intelligent agent who can emulate the behavior of a Blue (or if needed a Red) human being.

## Elements

In this section, the list of four high-level components of the approach is provided. The complete descriptions of these elements are provided in section Details of elements.

1. A web-based application
2. A cell phone-based application
3. Behavior learning software
4. Interaction learning software

The elements of the network are shown in yellow in Figure 1 (Figures are at the end of this paper). The definitions of the terms used in the figure are given in the previous section. The concept behind BlueNetwork is illustrated in Figure 2. The intra-connections among Red and Blue humans are harmonized and replaced by inter-connections between Red and Blue humans.

## Hypotheses

1. Social disorders and problems contribute to more IT, food and energy consumption.

   People suffering from these problems try to fulfill and answer to their needs by other means such as excessive movie watching, virtual life, etc. On another direction, these problems may affect the normal living schedule of person which results in sleep disorders and late night activities. All these parameters contribute in more IT consumption, more food consumption, and unhealthy easing patterns. These behaviors, directly and indirectly, result in extra energy consumption, carbon footprint, and water usage.

2. Obesity, although by itself is a global problem, can also be used as an indicator of a person's level of being at health risk (excluding athletics and healthy but obese people). Also, it is manageable in new generations (children) by healthy eating-and-living habits (expect for a low percentage who suffer from genetic obesity disease).

3. Friends and partners (social networks) have great impact on a person behavior, especially eating behavior. One of the most active areas of research into social relationships has been the study of network influences on positive and negative health behaviors (Kennedy2011). Engineering the social networks for better society can be considered as an effective tool (Cebrian2011).

4. Friends and partners (social networks) have great impact on obesity (Christakis2007, Fraser2002, Rhodes2005,Fletcher2011,Gest2011,Morris2011). Even, It has been stated that one person eats what his friends eat (Fletcher2011). Also, it has been observed that community-based interventions could be promising approaches, and also important components of a comprehensive response to obesity (Allender2011).

5. Communication (especially, non face-to-face communication) is the main means of interaction between friends (Hughes2003).

6. Obese people look for cure and treatment. Even, it has been observed that they are willing to pay for it (Fu2011).

7. In addition of direct costs of obesity, there are also some indirect psychological costs; an obese person may experience a reduced quality of life, for example through reduced ability to travel, reduced social networking, social marginalization (Apolloni2011), and reduced self-esteem. Studies show that overweight young women tend to have a more limited friendship circle, fewer years of school education, a lower likelihood of marriage, reduced employment prospects and reduced household income (Lobstein2011). All these factors contribute to higher living costs for the person.

## Hypotheses Summaries

Based on the hypotheses, a series of summaries is made here. They are the foundation of the 4-element approach we propose in the form of BlueNetwork. The elements will be discussed in Details of elements section.

1. Obesity, not only a global threat to the humanity, contributes to global warming.

2. Obesity is controllable in new generations by advertizing healthy behavior such as healthy eating habits (for example, fruit-based diet).

3. Obesity is measurable, and therefore is a practical score to build a model on top it.

4. Social networks and human interactions can be used to reduce obesity.

5. IT-based communications is a well-defined and low-cost interaction means.

# Details of elements

## A web-based application

This interface will provide a human-friendly connection between Red and Blue humans (and Purple agents in future). The communication between humans is selective, i.e., a Red human can only connect to a Blue one (not to other Red humans). People are asked to participate in the network as Blue or Red humans based on their own judgment of their obesity (supported by their BMI score). All members of network are encouraged to report two scores (weight and happiness) which will be used by the person connected to them to follow up with their changes. It is recommended to provide these two scores in a regular basis (daily or weekly updating is recommended). The weight will be used to measure the changes in the obesity of the person (using BMI and other measures), and the second one, happiness, will be used in analysis of communication between Blue and Red humans in order to extract effective communication segments (Element 4) which will be used finally in the design of Purple agents. Anonymous practice will be recommended to members.

In addition to the scores, Blue humans are encouraged to report (twit) their general daily schedule which not only provides a good example to their connected Red humans, it also enables us to collect this schedules and extract the best practices in Blue human behavior. The collected data will be learned by the behavior learning software (element 3), which will also provide a mean to regularize the Blue humans schedule in the form of providing assistance and secretary services to them.

Integration of BlueNetwork with other social networks (such as Facebook) will be considered. This not only makes it easier for the members to access to the services, it also can help them in fighting back with social network pressure they may feel in those other networks.

The Blue humans voluntarily participate in the program. It can be suggested that this activity is considered as a tax-deductable volunteer charity act.

## A cell phone-based application

Although abstract functionality of this element is all the same as element 1, it provides an easy access and interaction via other platforms such as personal portable and smart devices (cell phones and smart phones). This will reduce the interaction delay, and therefore, it will assumably increase the influence on Red humans.

## A Behavior (Schedule) learning software

Although the blue humans are considered well behaving, there is a spectrum of behavior in this group, and even for each individual blue human, his behavior is variable over time; and, depending on his environment and social interactions, his behavior may change. As Blue humans are the core and source of healthy influence in our network, protecting them and also adjusting their behavior is a key element for sustainability of the network. To achieve this, positive feedback for Blue humans are considered in the system in all interfaces of the network. This feedback is presented in the form of a secretary service to Blue humans. The secretary of a Blue human helps and reminds him with his schedule. Limiting the influence to the schedule has the benefit of having high-level and structural influence on the person.

Behind this secretary service, which works for keeping up healthy schedule of Blue humans, there is behavior learning software which learns the best behavior of each Blue human, and also the best behavior of their classes or groups. The software receives the information from the Blue human, and extracts regular behaviors and map them on their happiness score in order to find good schedules. Various cycles (weekly, monthly, seasonal, etc) should be considered in the analysis.

In order to discover the personal schedule of Blue humans, we will use ontology-based context-aware approaches (Zhang2011). In these approaches, the behavior learning model can provide feedback to the person based on context information such as current activity, location, time and personal schedule.

## An interaction learning software

Although the main goal of BlueNetwork is providing a directed and selective communication between Blue humans and Red humans in order to modify the behavior of Red humans and reduce their non-green footprints, high volume of specialized communications collected by BlueNetwork between Blue and Red humans provide a rich source of data for performing many analysis and research on human behavior, social networks and behavior engineering. One of immediate analyses is the understanding of the signals Blue humans send to Red humans in order to help them, increase their happiness, and make them more active. An interaction and behavior learning software will be developed to collect all the communications and analyze sequence of message pockets sent back and forth between Blue and Red humans in order to learn good sequences which improve the happiness the Red humans and decrease their obesity. The result of analysis will be used to design Purple agents which try to emulate good communication with Red humans. This not only makes the network less dependent on Blue humans and more self-sufficient and sustainable, it also provides a new level of human-machine interaction. In a very rare situation, the Purple agent could also emulate Red humans behavior for Blue humans; this could happen in the case of shortage of Red humans (which is actually the ultimate goal of BlueNetwork). In this case, the network could provide an effective environment for Blue humans in order to enable them to work with imaginary in-need people (Purple agents), and fulfill their psychological needs of helping and contributing to other human beings.

We will apply some linguistic methods to the collected communication sequences and segments, which can be considered as social linguistic actions (Kuo2011). The same technique can be used for studying the Internet as a linguistic communication system that hosts social networks, interactive sharing, dynamic collaboration, and social actions. In support of main idea behind BlueNetwork it can be stated that the speech acts (in the digital forms) are playing a key role in directing actions that affect the economic, political, or physical outcome of social actions.

## Outcomes

Immediate outcomes of the implementation of BlueNetwork (considering its pilot project) can be summarized as follows:

1. A communication tool (implemented on different platforms: web and cellphone) for directed and selective exchanges between humans toward lower obesity rate, better health and greener society.
2. A valuable collection of communication records between humans, especially conducted for improving their life style.
   a. As a sub-collection, a collection of Blue humans behavior in the form of their schedule.
3. An understanding model (and software) of the communications based on the data collected in item 2.
4. An understanding model (and software) of the healthy behavior (schedule) based on the data collected in item 2.a.

## Future prospects

Although obesity score is used in this network to differentiate Blue and Red humans, generalization to include other criterions (such as knowledge wealth, activity level, and so on) will be considered in future.

## Appendix: Obesity statistics and influence factors

In this appendix, some discussions on the global trend of obesity and social factors which influence the obesity rate are presented.

1. Obesity is a threat at the global scale. The obesity rates around the world are increasing (see Figure 3). Although the obesity rate depends on many life style parameters, the rates are in general higher compared to beginning of 90s.

2. As shown in Figure 4, it is predicted that by 2015 more than 2 billion will be overweight (more than 500 millions obese) which is a global alert (Booth2000). Food consumption of obese people by itself requires a large population of domesticated animals (cattle) which are considered a big source of GHG emissions. The global distribution of obesity is shown in Figure 5. Despite some fluctuations around the world, there is a global trend toward higher obesity rates. Another alert regarding obesity is high rates of obesity in children (Figure 6).

3. As shown in Figure 7, obesity has a big impact on economy (Adams2008). In addition to the food production costs, health-related costs, associated social-disorders costs, and transportation costs should be considered.

4. Obesity depends on the cultural background and social status (Sassi2009). Although obesity increases globally (Figure 8), the rates are not the same all over the world. Some cultures achieve lower rates that can be linked to social and cultural factors. The other influence factor is gender, and there is a big difference regarding the gender (Figure 9). This can be related to the gender-related social phenomena and interactions. It is interesting that in USA, which hosts the highest obesity rates, the overweight education level is the lowest compared to other developed countries (Figure 10). Also, there are some ethnic relations to obesity (Figure 11); the rate is higher for non-Hispanic black women. Finally, it seems that low-level of income also contributes to high obesity rate (Figure 12). This could be related to the lower level of education and social network of a child in low-income family. All these results show that there is a strong social influence in high obesity rates. Therefore, it is possible to propose a solution based on social networking in order to reduce the obesity rates.

# References

[1] Adams, R. J., Tucker, G., Hugo, G., Hill, C. L., & Wilson, D. H. (2008). Projected future trends of hospital service use for selected obesity-related conditions. *Obesity Research & Clinical Practice*, *2*, 133–141.

[2] Allender, S., Nichols, M., Foulkes, C., Reynolds, R., Waters, E., King, L., Gill, T., Armstrong, R., & Swinburn, B. (2011). The development of a network for community-based obesity prevention: the CO-OPS collaboration. *BMC Public Health*, *11*, 132.

[3] Apolloni, A., Marathe, A., & Pan, Z. (2011). A longitudinal view of the relationship between social marginalization and obesity. In J. Salerno, S. Yang, D. Nau, & S.-K. Chai (Eds.), *Lecture Notes in Computer Science: Social Computing, Behavioral-Cultural Modeling and Prediction (SBP'11)* (pp. 61–68). Springer Berlin / Heidelberg volume 6589.

[4] Booth, F. W., Gordon, S. E., Carlson, C. J., & Hamilton, M. T. (2000). Waging war on modern chronic diseases: primary prevention through exercise biology. *Journal of Applied Physiology*, *88*, 774–787.

[5] Cebrian, M., & Pentland, A. (2011). Engineering trade-offs in social organization: The beginnings of a computational social science. *IEEE Instrumentation & Measurement Magazine*, *14*, 18–21.

[6] Christakis, N. A., & Fowler, J. H. (2007). The spread of obesity in a large social network over 32 years. *New England Journal of Medicine*, *357*, 370–379.

[7] Fletcher, A., Bonell, C., & Sorhaindo, A. (2011). You are what your friends eat: systematic review of social network analyses of young people's eating behaviours and bodyweight. *Journal of Epidemiology and Community Health*, *65*, 548–555.

[8] Fraser, S. N., & Spink, K. S. (2002). Examining the role of social support and group cohesion in exercise compliance. *Journal of Behavioral Medicine*, *25*, 233–249.

[9] Fu, T.-T., Lin, Y.-M., & Huang, C. L. (2011). Willingness to pay for obesity prevention. *Economics & Human Biology*, *9*, 316–324.

[10] Gest, S., Osgood, D., Feinberg, M., Bierman, K., & Moody, J. (2011). Strengthening prevention program theories and evaluations: Contributions from social network analysis. *Prevention Science*, *Online first*, 1–12.

[11] Hughes, S. (2003). The use of non face-to-face communication to enhance preventive strategies. *Journal of Cardiovascular Nursing*, *18*, 267–273.

[12] Kennedy, D. P., Green, H. D., McCarty, C., & Tucker, J. S. (2011). Non-experts' recognition of structure in personal network data. *Field Methods*, *23*, 287–306.

[13] Kuo, F.-Y., & Yin, C.-P. (2011). A linguistic analysis of group support systems interactions for uncovering social realities of organizations. *ACM Trans. Manage. Inf. Syst.*, *2*, 1–21.

[14] Lobstein, T. (2011). Prevalence and costs of obesity. *Medicine*, *39*, 11–13.

[15] Morris, M. E., Consolvo, S., Munson, S., Patrick, K., Tsai, J., & Kramer, A. D. (2011). Facebook for health: opportunities and challenges for driving behavior change. In *Proceedings of the 2011 annual conference extended abstracts on Human factors in computing systems* (pp. 443–446). Vancouver, BC, Canada: ACM.

[16] Sassi, F., Devaux, M., Cecchini, M., & Rusticelli, E. (2009). *The Obesity Epidemic: Analysis of Past and Projected Future Trends in Selected OECD Countries*. Technical Report 45 OECD Health Working Papers.

# The BlueNetwork Concept (Figures)

***Reza FARRAHI MOGHADDAM, Fereydoun FARRAHI MOGHADDAM and Mohamed CHERIET***

Synchromedia Lab, ETS, University of Quebec
1100 Notre-Dame west, Montreal, H3C 1K3 Canada

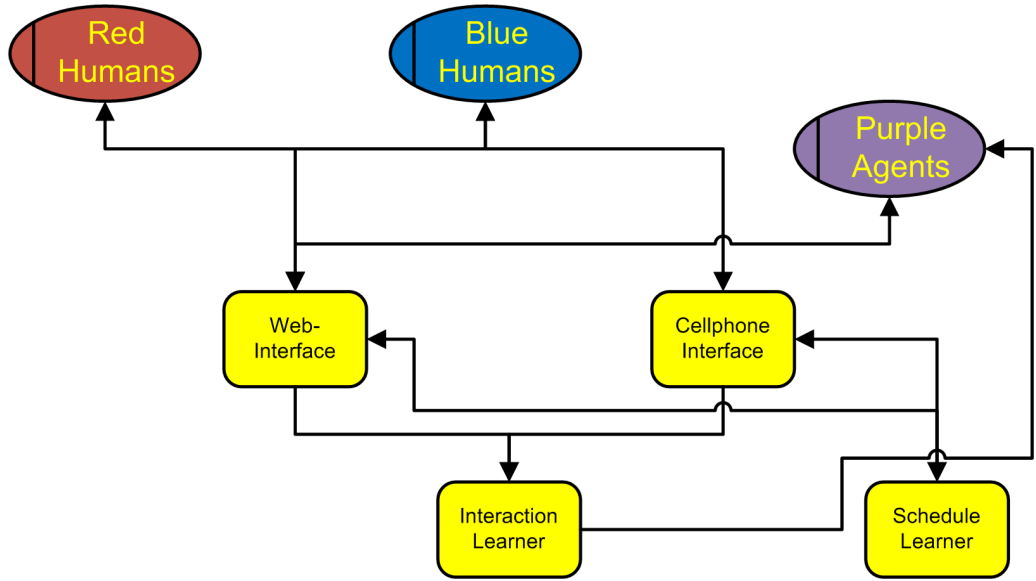

Figure 1. The elements of BlueNetwork.

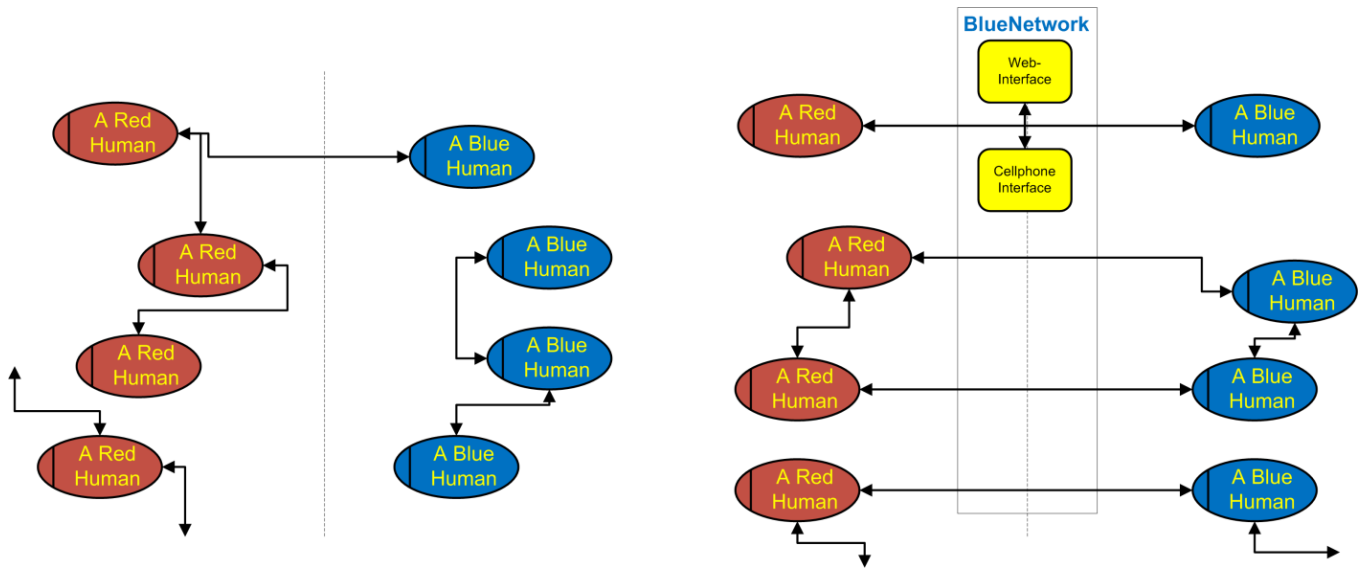

Figure 2. The BlueNetwork concept. Left: Connections without BlueNetwork. Right: Organized connection using BlueNetwork.

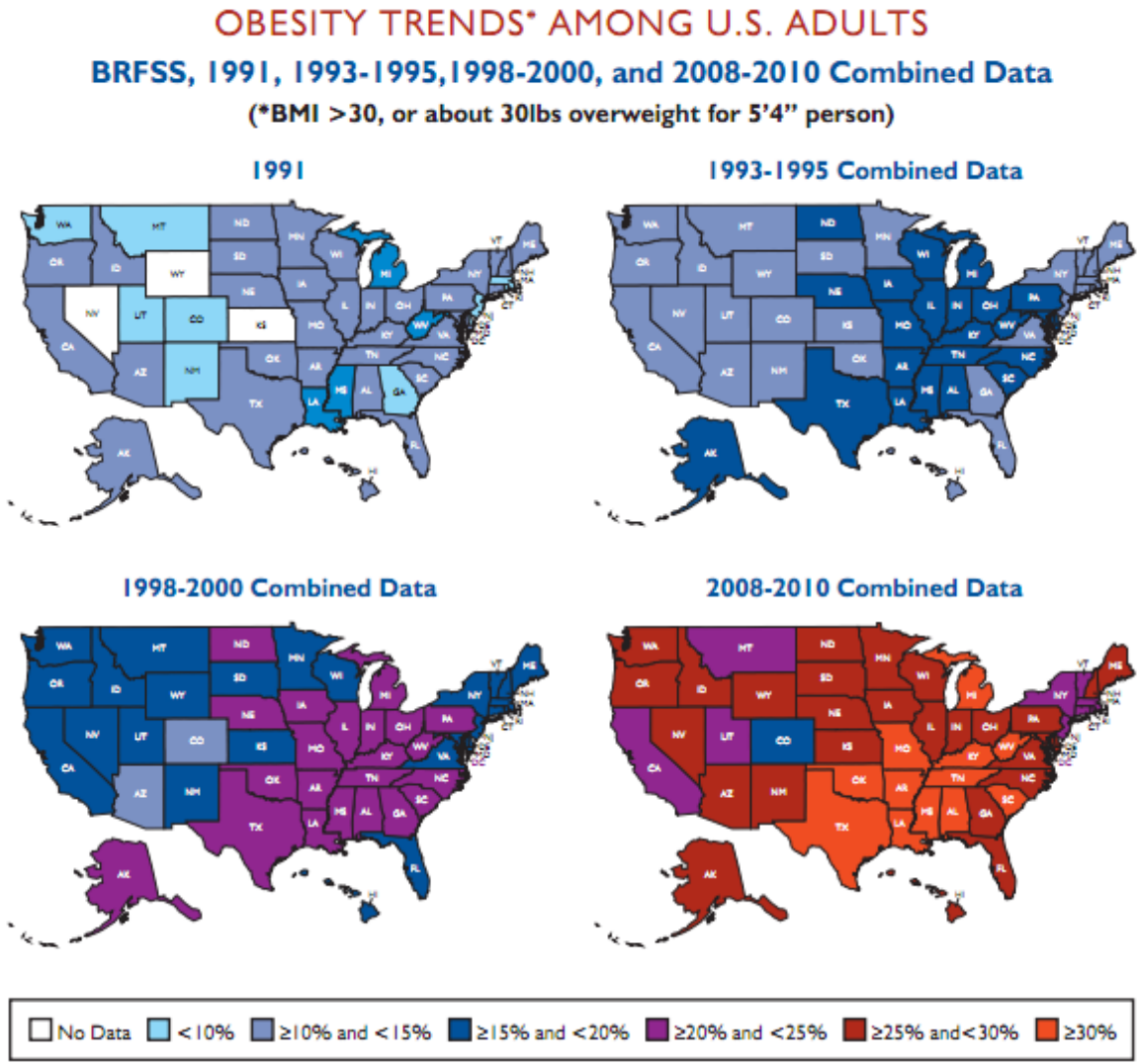

Figure 3. Obesity trends among US adults.

Source: http://paul.kedrosky.com/archives/2011/07/obesity-rates-in-u-s-only-colorado-left-under-20.html

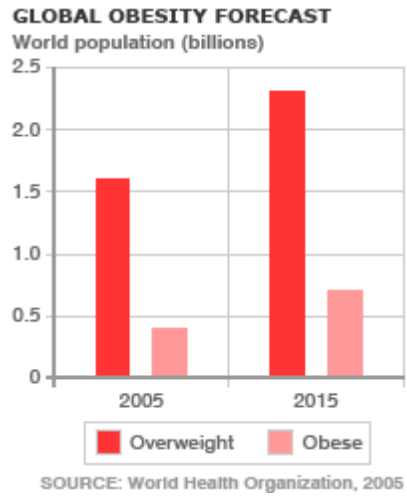

Figure 4. Obesity rate projection for 2015.

Source: http://nur261alanpjack.blogspot.com/2008/05/university-of-plymouth-faculty-of.html

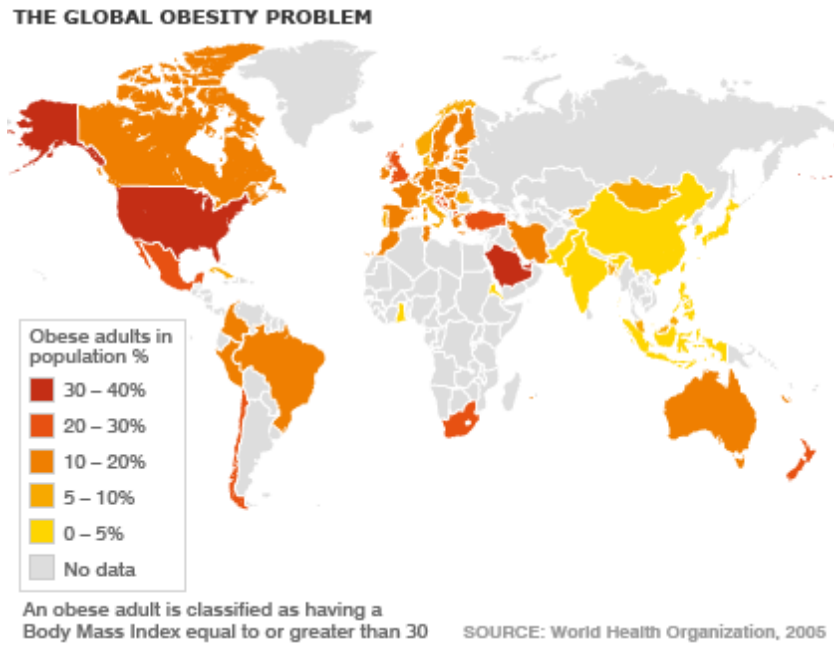

Figure 5. Global distribution of obesity.

Source: http://nur261alanpjack.blogspot.com/2008/05/university-of-plymouth-faculty-of.html

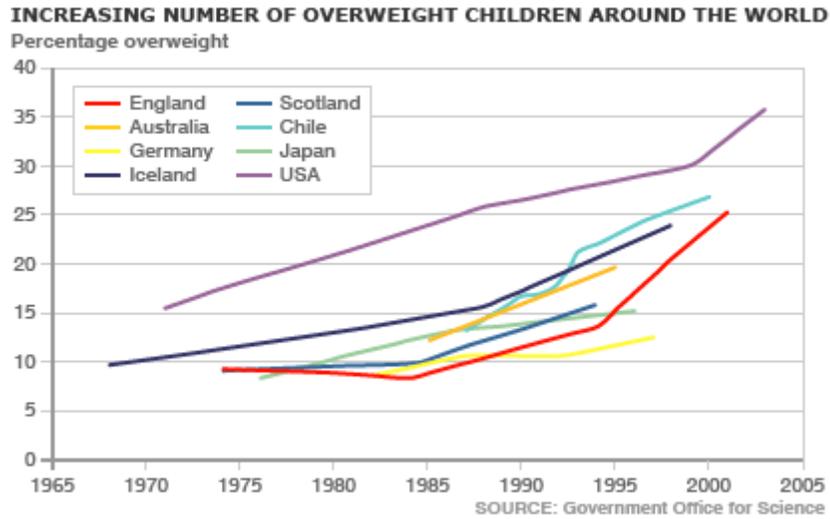

**Figure 6. Obesity rates around the word.**

Source: http://nur261alanpjack.blogspot.com/2008/05/university-of-plymouth-faculty-of.html

| Inactivity is a major public health threat | | | | |
|---|---|---|---|---|
| | **Alcohol misuse** | **Smoking** | **Obesity** | **Inactivity** |
| Percentage of adult population affected in England | 6–9% | 20% | 24% | 61–71% |
| Estimated cost to the English economy per year | £20 billion | £5.2 billion | £15.8 billion | £8.3 billion |
| Estimated cost to the NHS per year | £2.7 billion | £2.7 billion | £4.2 billion | £1–1.8 billion |
| Source: Adapted from a number of reports of official statistics and other published work | | | | |

**Figure 7. health costs associated with obesity.**

Source: http://dailymonthly.com/?p=363

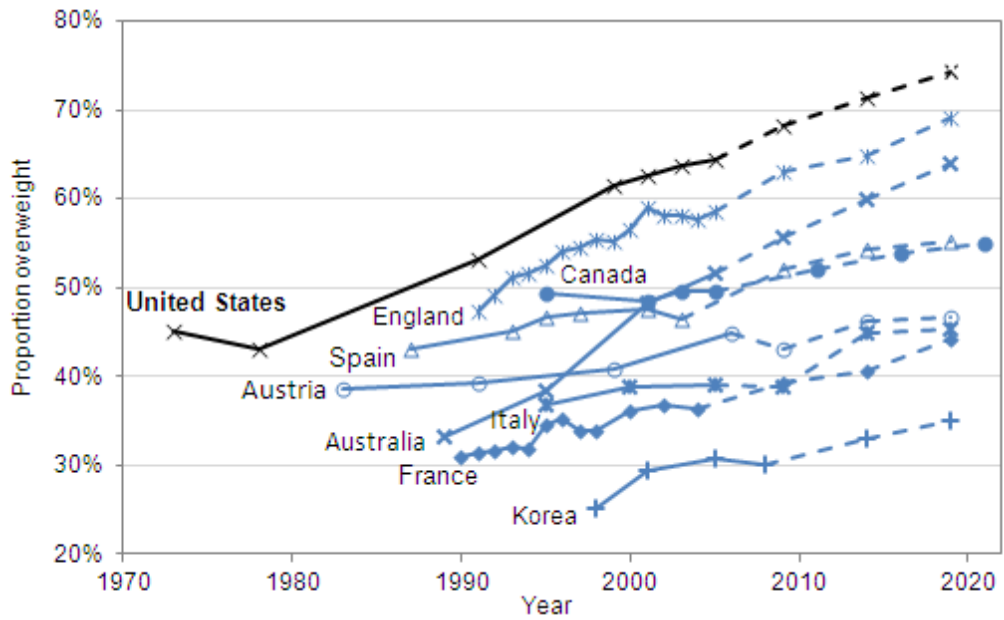

Figure 8. Overweight rates in different countries.

Source: http://www.oecd.org/document/57/0,3343,en_2649_33929_46038969_1_1_1_1,00.html

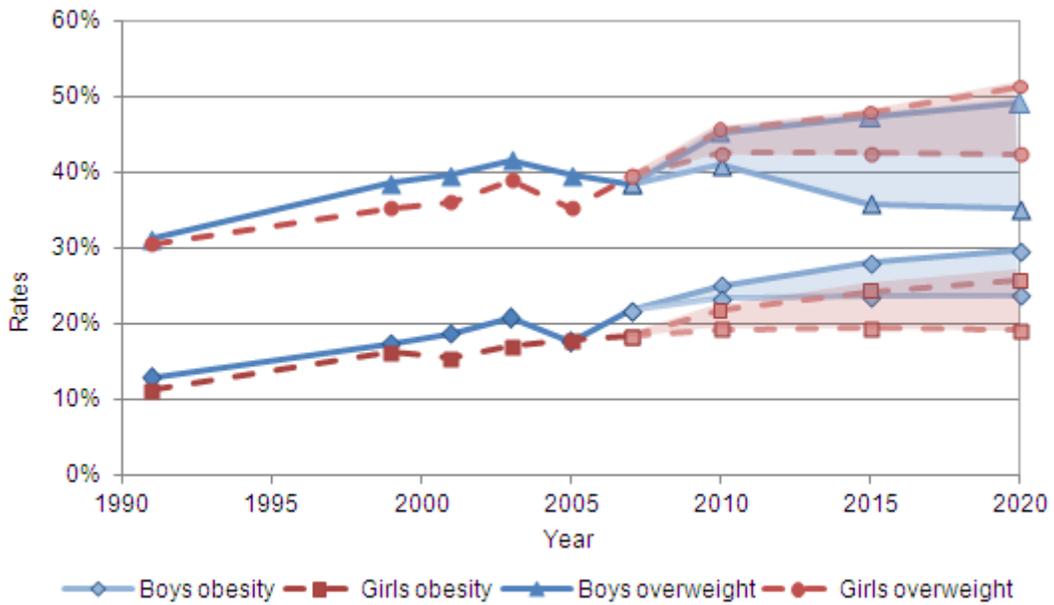

Figure 9. Obesity rates among boys and girls.

Source: http://www.oecd.org/document/57/0,3343,en_2649_33929_46038969_1_1_1_1,00.html

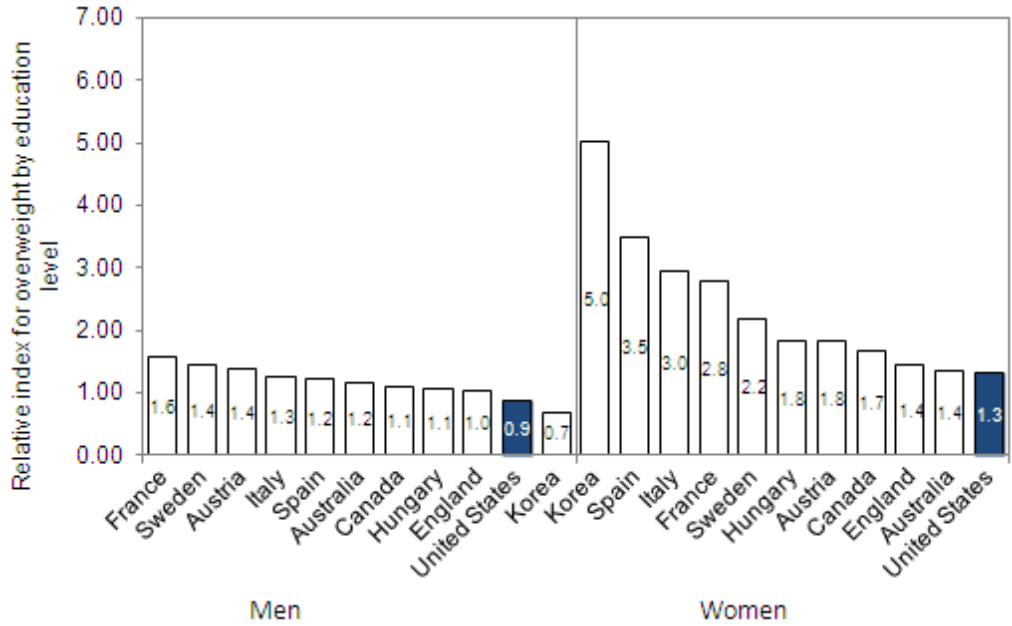

Figure 10. Overweight education around the world.

Source: http://www.oecd.org/document/57/0,3343,en_2649_33929_46038969_1_1_1_1,00.html

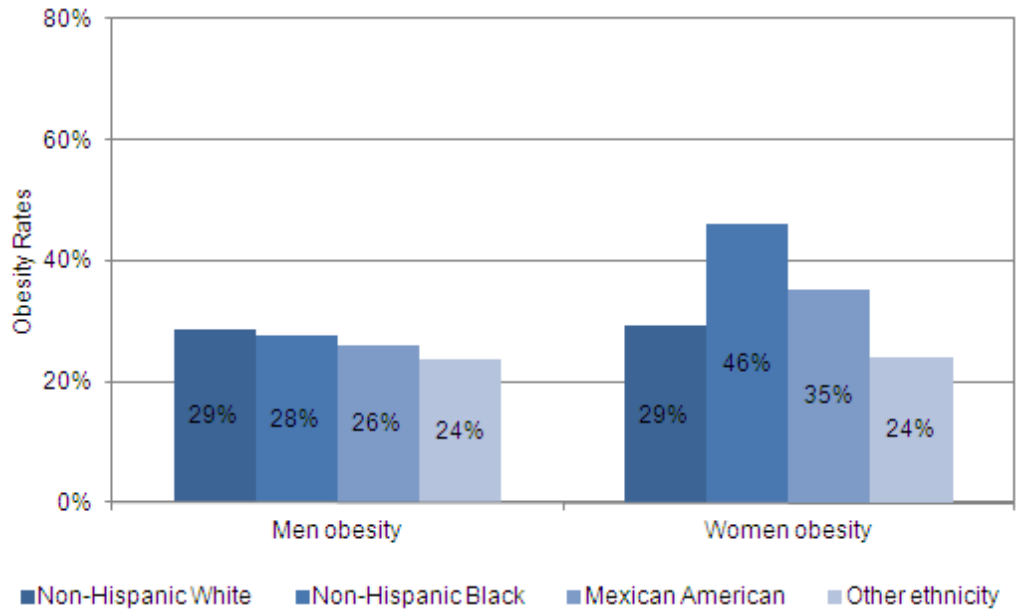

Figure 11. Obesity rates versus ethnicity.

Source: http://www.oecd.org/document/57/0,3343,en_2649_33929_46038969_1_1_1_1,00.html

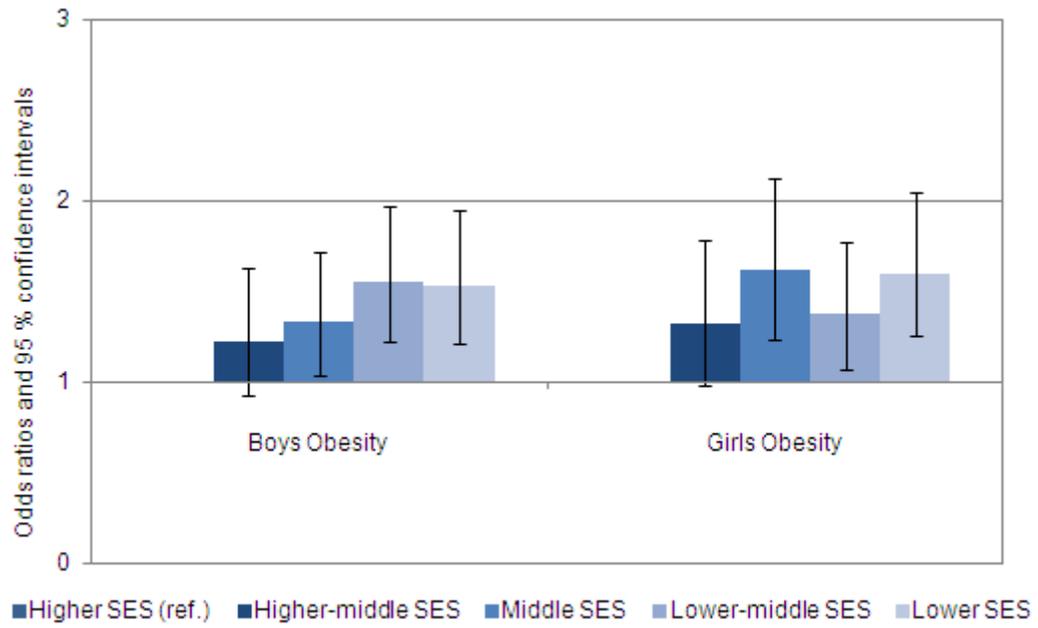

Figure 12. Obesity rates versus income.

Source: http://www.oecd.org/document/57/0,3343,en_2649_33929_46038969_1_1_1_1,00.html